\documentclass[b5paper]{article}
\usepackage{amssymb}
\usepackage{graphicx}
\usepackage{longtable}

\begin{document}

\title{Identification and astrometry of variables in M3 }
\author{by \\
G\'asp\'ar \'A. Bakos\thanks{e-mail:~{\tt bakos@konkoly.hu}} 
\thanks{Corresponding WWW: {\tt http://www.konkoly.hu/staff/bakos/M3/}},
J\'ozsef M.
Benk\H{o} and Johanna
Jurcsik \\
{\normalsize Konkoly Observatory, P.~O.~Box~67, H--1525 Budapest, Hungary}}
\maketitle

\begin{abstract}
We present identification and astrometry of all previously known or
suspected variables along with the discovery of six new variables in
the globular cluster M3.  The number of the catalogized variables
increased to 274 by including all the confirmed, previously known
variables and the new discoveries.  The precise and homogeneous
astrometry, as well as the clarification of misapprehensions in the
preceding identifications are done by using overlapping fields from a
wide-field Schmidt-camera, a 1m RCC telescope, and HST archive
observations from the center of the cluster. The astrometric positions
can serve as a direct input to any photometry which needs the accurate
centers of the variables.
\end{abstract}

\section{Introduction}

Starting with the discovery of a W Vir star in M3
\cite{Pickering1889}, which was the first pulsating variable to be
observed in any globular cluster, M3 deserved the reputation of being
the richest globular cluster in variables. The pioneering work of
Bailey at the turn of the century yielded more than hundred variables,
mostly RR~Lyrae stars \cite{Bailey1902}.  By 1939, following
discoveries of Shapley, Larink and M\"uller increased the number of
catalogized variables in this cluster to 201, though variability of
several candidates was later refuted, and misidentifications were
prevalent \cite[and references therein]{SH39}. Observations of
Sandage, Kurochkin, Kukarkin, Kholopov, Russev, Meinunger and Kaluzny
led to the extension of later editions of the Variable Stars in
Globular Clusters to 225 variables for M3 \cite[and references
therein]{SH55,SH73}, and recently to 238 entries \cite{CC98}.

However, erroneous identifications are present, partly because the
coordinates respect to the cluster's (not uniformly accepted) central
position were taken from various sources, partly due to the severe
crowding conditions. Moreover, in some cases astrometry was carried out
on photographic plates exposed under imperfect seeing conditions, and
by manually centering on the stars' profiles. \cite{ESTS94} compiled a
precise list of positions for all known or suspected variables using a
{\em homogeneous} reference grid. As several new variable discoveries
have sprung up since 1994, compilation of a new astrometric list with
cross references became substantial.

This work is based on an observing campaign of M3 with a 0.9m Schmidt
and a 1m RCC telescope, consisting of several observing runs in the
spring seasons of 1998 and 1999. We primarily lay emphasis on
identification and astrometry, but variability of the candidates is
also confirmed in numerous cases. Photometric aspects and results on 
individual variables will be dealt with in forthcoming papers
\cite{Benko2000, Bakos2000}.

\section{Observations}

As concerns the peripheral variables, we used CCD observations obtained
with the 60/90/180cm Schmidt telescope at the Piszk\'estet\H{o}
Mountain Station of Konkoly Observatory. Observations in V-band were
carried out on 11 nights evenly scattered in three months of the the
spring season of 1998. The Kodak KAF-1600 $1024\times1536$ chip yielded
a $21'\times28'$ field of view (FOV), with
$\sim~\!1''/\rm pixel$ resolution. Only the best seeing
($\sim1.5''$) image was used for astrometry, but all of them were
employed so as to verify the variability of the sources.

While identification of outer variables is straightforward in
principle, crowding conditions impose difficulties in the inner few
arcminutes. A subset of observations carried out with the 1m, f/13.6
RCC telescope at the same observatory was used to patch astrometry in
the dense, inner region. The TH7896M UV-coated $1024\times1024$ CCD chip
attached to the telescope yielded a FOV of
$\sim5'\times5'$, with $0.288''\rm pixel$ resolution.
The observing run in March 1999 produced time-series data for 5 nights
with strongly varying seeing conditions (ranging from $1.2''$ to
$\gtrsim 2.5''$). Only the best night, and only V-filter observations
were used in the present work.

As the innermost region of M3 was still not clearly resolved, we used
archive Hubble Space Telescope (HST) Wide Field Planetary Camera 2
(WFPC2) observations\footnote{ Based on observations made with the
NASA/ESA Hubble Space Telescope, obtained from the data archive at the
Space Telescope Science Institute. STScI is operated by the Association
of Universities for Research in Astronomy, Inc. under NASA contract NAS
5-26555.} of M3 (proposed by Fusi Pecci, 1994, cf.~Ferraro et
al.~1997). This is the only set of publicly available \textsc{WFPC2}
observations where frames have short enough exposure times ensuring
that RR~Lyrae variables are not saturated. We used an F814W mosaic
image with 3s integration time designated as ``u2li010dt'' in the
archive.

The standard \textsc{iraf/ccdred} package was used for overscan and
flatfield correction of the Schmidt and RCC images. The HST archive
image was already calibrated by the HST pipeline, and while we were not
performing direct photometry on the image, recalibration was needless.

\section{Astrometry}

All astrometric work was  based on the USNO-A2.0 catalogue, which had
been derived from PMM (Precision Measuring Machine) scans of POSS-I O
and E plates \cite{Monet1996,Monet1998}. The typical astrometric error
is about $0.15''$ for individual stars away from the plate corners
and not in crowded fields. After thorough examination of crowding as a
function of radial distance from the center of M3, we selected
USNO-A2.0 stars with radial distance $r>6'$ and brighter than
$M_{red}=17.8^m$ (hereafter {\em grid0}). Uncertainties in USNO
astrometry of stars closer to the center might have led to considerable
errors, ie.~it was not possible to use USNO stars for reference in the
RCC and HST fields. By careful choice of overlapping regions on the
Schmidt, RCC and HST frames, the astrometric reference grid was
extended towards the center.  Equatorial coordinates given in the
International Celestial Reference Frame (ICRF) were transformed to
relative coordinates ($\Delta\alpha$ in $15\times \rm
seconds$\footnote{This unit was chosen for convenience, as it has the
same order of magnitude as arcseconds for small declinations. Values in
Table~1 are given in {\em normal second} units.},
$\Delta\delta$ in arcseconds) respect to the center\footnote{Note that this
center slightly differs from the one accepted in \cite{CC98} precessed
for J2000.0} of M3 (J2000.0):
$\alpha_{M3}=13^{\rm h}42^{\rm m}11.2^{\rm s}$,
$\delta_{M3}=28^{\circ}22'32''$ \cite{Harris1996}.

Using our self developed \textsc{iraf/gastro} astrometric
package\footnote{All self-written software are available from the first
author via request in e-mail.}, we selected isolated, not saturated
stars with $r>6'$ on the Schmidt image ({\em grid1a}). The
centers of these sources were determined by a script
(\textsc{gast\_mkgrid}) built up from the \textsc{iraf/daofind, phot,
pstselect} programs.

Using a best estimate of the transformation between {\em grid0} and
{\em grid1a} (initially a manual guess by \textsc{gast\_crude}), stars
were cross-identified, and transformations were refined. This iteration
was done several times with variable rejection thresholds and fitting
parameters, see-sawing back and forth between identification and
refinement (\textsc{gast\_prec}). Finally we fitted a second-order 
transformation between {\em grid0} (ICRF) and {\em grid1a} (Schmidt)
of the form
\begin{eqnarray*}
X_{ICRF} & = &
A_{11}+A_{21}X_{Sch}+A_{12}Y_{Sch}+A_{31}X_{Sch}^2+A_{22}X_{Sch}Y_{Sch}+A_{13}Y_{Sch}^2\\
Y_{ICRF} & = &
B_{11}+B_{21}X_{Sch}+B_{12}Y_{Sch}+B_{31}X_{Sch}^2+B_{22}X_{Sch}Y_{Sch}+B_{13}Y_{Sch}^2,
\end{eqnarray*}
which adequately conforms to the large-scale distortion of the
Schmidt-camera field. The typical scatter of least-squares fits based
on $\sim150$ evenly scattered stars was lower than $0.13''$ rms,
which is even better than the USNO-A2.0 precision itself. 

As USNO-A2.0 astrometry is not reliable in RCC fields covering the
dense center of the cluster ($5'\times5'$), we extended our
Schmidt grid to the innermost $3'>r>1.5'$ ring ({\em
grid1b}), and in a similar manner, established a grid on the RCC fields
with $r>1.5'$ ({\em grid2a}). The transformation between the two
grids, based on 250 common stars, was derived similarly as previously
mentioned, and had an overall scatter less than $0.04''$.

Finally, transformations were calculated between the RCC images ({\em
grid2b:} $r<1.5'$) and the HST PC, WF2, WF3, WF4 chips, using
70, 150, 150 and 200 stars, respectively. Scatter in the transformations was
less than $0.04''$.

On the whole, it became possible to transform any astrometric position
from the HST, RCC and Schmidt images to equatorial system (ICRF) with
$\lesssim 0.15''$ precision.

\section{Identification of variables}
\subsection{Tools}

Four basic tools were applied to identify variables: coordinate lists,
finding charts, variability images and light-curves. Relative
coordinates for 238 variables are given in \cite{CC98}. Unfortunately
these coordinates are collected from several sources, thus far from
being homogeneous, and the precision is only $\pm0.1\--1.0''$,
which is not eligible for unambiguous identification in the center.
Relative coordinates in right ascension are (traditionally) given in
arcseconds as opposed to seconds, which involves a projection of the
spherical equatorial coordinate system. \cite{ESTS94} give a more
accurate list with proper equatorial coordinates. To be on the safe
side, we used both lists to filter out misidentifications. New HST
discoveries from \cite{GU94}, properly transformed to equatorial system
by \cite{Goranskij1994} were also used.  Finding charts are enclosed only 
in a few sources, such as \cite{Kholopov1963,Kholopov1977} and
\cite{Kaluzny1998}.

As concerns the variability images, we applied the Image Subtraction
Method (ISM) for the best night on the RCC frames, and for all 11
nights on the Schmidt frames, as described in \cite{Alard1998} and
\cite{Alard1999}. After registering and convolving the images, then
subtracting them from the reference frame with the aid of
\textsc{isis2.0}, the resultant difference images were almost blank,
only variable stars having negative or positive profiles. The
variability image \cite{Olech1999} was the average of the absolute
values of several difference frames, containing accumulated
contributions from all variations respect to the reference frame, that
is showing all sources variable on the timescale of the observations
($\sim 40$ and $25$ images for the Schmidt and RCC observations,
respectively). In the case of the RCC observations, we restricted our
investigation to variability detectable in one single night (mostly RR
Lyraes and perhaps SX Phe stars), data reduction of subsequent frames
will improve identification of both longer period (RGB) and fainter
variables. On the other hand, the Schmidt observations spanning three
months were also convenient for detecting long period irregular
variables. 
It has to be emphasized that the primary goal was
identification of the known variables and suspected ones, 
and not a complete variable
search, although six new variables were found.  
In some complex cases
(mostly close doubles) we also made use of scrutinizing the dependence
of the light-curves on the (fixed) center of the photometry; the
smaller scatter of the curve implies that the center of photometry is
closer to the real position of the variable, which can serve as a
criterion for selecting the truly varying star.

We implemented our \textsc{iraf/gisis/gvarfind} task, which blinks the
normal and the variability images, thus identification can be judged
not only upon the initial coordinate-estimate but also on the
variability of the source.  This method seemed to be extremely useful
in the central areas, where sometimes several candidates were
equidistant from the corresponding rough position. The aforementioned
task also performed fitting of an elliptical gauss profile to the
selected source, either on the normal image or on the variability
image. If variability was sufficient, the latter case was preferred, as
non-variable neighbors completely disappeared even in the dense
regions, where profile-fitting would have been biased by overlapping
profiles. It is worthy to note that HST variability image was not
available, as we had only a single image with not saturated RR Lyrae
stars.

In all cases identification meant determining the position of the
variable by {\em precise psf-fitting}, and adequately {\em
transforming} the coordinates to the ICRF.

Photometric accuracy is an important aspect when confirming variability
or invariance of a star. Investigation of light curves derived by
aperture photometry on the Schmidt and RCC frames yielded a crude
estimate of the photometric errors being smaller than $\delta V\lesssim
0.05^m$ and $\delta V\lesssim 0.03^m$, respectively. However,
significantly smaller accuracy could have been achieved in the most
crowded fields.

As the Image Subtraction Method is less sensitive to crowding
conditions, the accuracy of the (ISM) light curves we checked for
variability were in the order of hundreds of mags in all cases.
Variability of each star was checked and commented only if no
light-curve had been previously given.

\subsection{Finder charts}

We present finder charts for 286 variables and suspected
variables within the boundaries of the RCC frame (Figs.~1 to 10), or
within the Schmidt frame (Figs.~11 to 15), to avoid further
misidentifications and assist in any variability study of M3. 
Variables on the very edges of the Schmidt frame were omitted (V17,
V92, V117), as well as variables denoted by ``E'' in Table~1 (V82, V91,
V112, V113, V114, V115, V123, V141, V205, V206, V230). Although
variable vZ1283 was identified on the edge of the RCC frame, the finder
chart is given on the Schmidt image. Two HST variables (GU9016, GU9025)
were not included as they were not readily visible on the RCC images
(both identified from the HST frame). Several candidates revealed as
doubles on the HST images were not resolved on the RCC frame, in which
cases only one chart was given (V159a,b, V192a,b, KG8a,b, X14a,b).
Though good quality finder charts for V237, V238 has been recently
published by \cite{Kaluzny1998} they are repeated here for
completeness.

\paragraph{Figures 1.-10}
Caption: Finding charts for variables on the RCC frames. Boxes are
approximately $15''$ in width, North is up, East is to the left. Boxes
from very sparse regions labeled as ``WF'' are $30''$ wide.  Different
intensity scales are choosen to reach the clearest resolution of each
frame. (Note: figures are enclosed in png format for compression)

\paragraph{Figures 11.-15.}
Finding charts for variables on the Schmidt frames. Boxes are
approximately $80''$ in width, North is up, East is to the left.
(Note: figures are enclosed in png format for compression)\\

The revised positions and some additional information on the variables
are given in Table~1. The IDs of the variables were extended in a
similar manner to the \cite{CC98} catalogue in case of all confirmed
variables (column~1.). The last column gives all the crucial
cross-references. The type of variability indicated in the second
column of Tab.~1 was determined from our observations if light
variation was detected.  For some of the HST variables classification
given by \cite{GU94} was adopted. Columns 3--5 and 6 give the relative
coordinates and the observation used for the identification,
respectively. Any comments of the individual stars are indicated in
column~7.

Whenever possible, variables were identified on the HST image, as this
yielded the most precise astrometry due to the $0.05''$ and
$0.1''$ per pixel resolution for the PC and WF chips,
respectively. Furthermore, due to the enhanced resolution, several
variables turned out to be close double stars, in which cases original
notation was split (eg. 122a, 122b)\footnote{Notation was also
splitted in case more than one candidates were equidistant from the
position given in previous references, and identification was
dubious.}. As we could not check variability on the HST frames, we
tried to identify the variable component using the profile on the RCC
variability frame. If a star was not within the WFPC2 boundaries, RCC
images were checked, and in case the star was beyond these frames,
Schmidt images were used. In eleven cases when variables were even
outside the Schmidt frames, we accepted positions from \cite{ESTS94}.
Still, our grid of variables remained self-consistent, as our
coordinates are compatible with
\cite{ESTS94} in the sense that no discernible offset is present (see
later). 

Kholopov variables (X) were manually selected with \textsc{gvarfind}
using finder charts in \cite{Kholopov1963,Kholopov1977} and our
variability image. Some of the Kholopov variables were already included
in \cite{CC98}, without firm constraints on their light-curves. All
these cases were checked for variability.  We identified the remaining
11 suspected candidates, and confirmed variability of five of them.

\cite{KG80} variables (KG) were identified from the list in the
original paper, and from \cite{ESTS94}. 
Variability of several
candidates is confirmed, and new ID numbers are given.

All von Zeipel (vZ) candidates listed in \cite{ESTS94} were identified
on the RCC and Schmidt images. As all of them were suspected as
long-period variables \cite{Welty1985,Meinunger1980}, variability was
checked on the Schmidt frames. Variability is ascertained only in one
case (vZ297), while six other candidates remain suspected variables. 
The remaining stars are constant within the photometric errors of the
Schmidt measurements.

Sandage variables (S-) were identified with the aid of finder charts in
\cite{Sandage1953}. Except for S-AQ, S-I-II-52, S-I-II-54, these stars
were suspected as bright, long-period variables \cite{Meinunger1980},
however all, but one (S-I-VI-65) proved to be constant on the
Schmidt frames.
Bao-An et al. (1993a,b) and \cite{Bao-An1994} drew attention
to
three new short-period, low-amp\-li\-tu\-de variables (S-I-II-52, S-I-II-54,
S-AQ). These stars were identified from charts in \cite{Sandage1953},
but low amplitude variability has not yet been confirmed on the Schmidt
images.

\cite{GU94} gave a list of variable stars (GU) in the center of M3 using
\textsc{HST/WFPC} observations, giving unique ID numbers, relative
coordinates respect to star AC999, rough average magnitudes, and
classification.  The list was appropriately transformed by
\cite{Goranskij1994} to match \cite{ESTS94}, and cross-identification
showed that only 11 out of the 40 variables were new discoveries. By
performing the same analysis, we reduced the number of new HST
discoveries to 9 (confer revised cross-references in Table~1).
Variability is confirmed except for GU1711 (long-period suspected
variable) and GU9016 (faint SX~Phe star).

Three prominently variable RR~Lyrae stars were found on the RCC images,
and three long-period variables on the Schmidt images, without any
previous reference. These supposedly new variables are listed at the
end of Table~1.  Preliminary light-curves for the newly discovered
variables 269--274 are presented in Fig.~16.

\setcounter{figure}{15}
\begin{figure}
\vspace*{-2.5cm}
\centerline{\includegraphics[keepaspectratio,width=12cm]{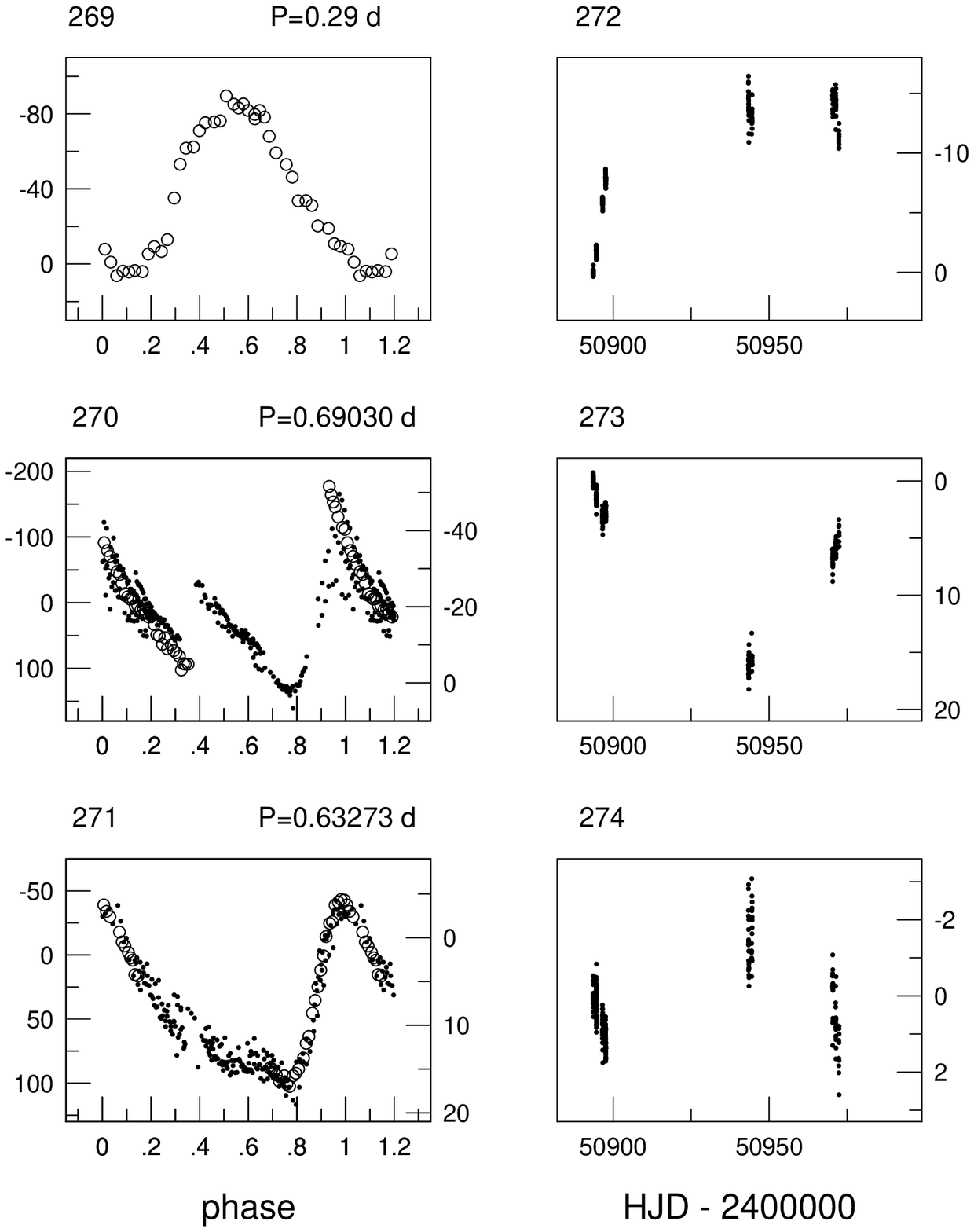}}   
\vspace*{-2.5cm}
\caption{
Light-curves obtained using image subtraction method (\textsc{isis2.0}) 
of the new variables.
The RR Lyraes (V269--V271) are located in the inner part of M3; both
the 1m RCC (denoted by open circles) 
and Schmidt observations (dots) are plotted. 
Scales (in units of differential fluxes (ADU/s) normalized with a
factor of 1000) are shown 
on the left and right axes, respectively. 
The large scatter and the jump in the light-curve of V270 are supposed 
to be defect caused by the inappropriate resolution of the Schmidt
images in a very dense region (see Fig.~9).
For long period variables (V272--V274) lying in 
the outer region of the cluster only Schmidt data was available.}
\end{figure}

Coordinates in the resultant list were correlated with the original
list of \cite{CC98}, and with the exception of few misidentifications
and defective coordinates in the above reference (V146, V151, V165,
V181, V196, V199, V204), the overall scatter of the fits was
$\sim1''\,\rm rms$. Correlation with the list of \cite{ESTS94} yielded
considerably more precise fits with $\lesssim0.2''\,\rm rms$. This
reinforces the {\em absolute} precision of our astrometry being better
than $0.2''$, even with taking the uncertainties rising from the
USNO-A2.0 catalogue into account. As concerns relative astrometry, our
grid is self-consistent within the error of $\sim0.15''$.

\section*{Acknowledgments}
We are grateful to Dr.~B\'ela Szeidl for his persevering help and expertise
in M3 variables. This project was supported by OTKA grants T-24022 and
T-30954. 

\eject
\renewcommand{\thefootnote}{\alph{footnote}}
\begin{longtable}{llrrrlll}
\caption{List of variables and suspected variables in M3$^{\rm a}$} \\
\hline\hline
\\[-7pt]
ID\footnotemark[2] & Type &  $\Delta\alpha(sec)$ &
$\Delta\alpha('')$\footnotemark[3] &
$\Delta\delta('')$ &
Tel.\footnotemark[4] & Comm.\footnotemark[5] &
Cross-ref.\footnotemark[6] \\[3pt]
\hline 
\endfirsthead
\caption[]{(continued)}\\
\hline\hline
\\[-7pt]
ID  & Type & $\Delta\alpha(sec)$ &
$\Delta\alpha('')$ & $\Delta\delta('')$ &
Tel. & Comm. &
Cross-ref. \\[3pt]
\hline
\\[-9pt]
\endhead
\\[-9pt]
\hline
\endfoot
\\[-9pt]
\hline\hline
\endlastfoot

\footnotetext[1]{Center for M3 is adopted from \cite{Harris1996}:
$\alpha_{M3}=13^{\rm h}42^{\rm m}11.2^{\rm s}$,
$\delta_{M3}=28^{\circ}22'32''$ (J2000.0)} 

\footnotetext[2]{Numeric identification 1 to 238 were listed in \cite{CC98}
and 239 to 274 is added in this work. If the star
turned out to be double on HST images, notation is splitted to ``a,b\dots'',
etc. The same notation is used if there are more than one candidates.}

\footnotetext[3]{This column is given for compatibility with previous
works, assuming a projection from seconds to arcseconds using the accepted
center of the cluster.}

\footnotetext[4]{Telescope used for identification: H \--- HST/WFPC2, R
\--- RCC, S \--- Schmidt, E \--- coordinates from \cite{ESTS94}}

\footnotetext[5]{Comments are: 
``c'' \--- extra comment on the star, 
``d'' \--- double star,
``e'' \--- on the edge of frame,
``f'', ``f?'' \--- variability confirmed and suspected
from corresponding RCC variability image 
and light curve,
``i'' \--- identification problems, 
``j'', ``j?'' \--- variability confirmed and suspected on Schmidt images,
``k'' \--- not variable on Schmidt images,
``m'' \--- merging with closeby star,
``n'' \--- variability not detected on RCC images.}

\footnotetext[6]{
 ``X'' stands for \cite{Kholopov1963,Kholopov1977}, ``KG'' for \cite{KG80},
``GU'' for \cite{GU94}, ``S'' for \cite{Sandage1953}, ``Sh'' for
\cite{Shapley1914}, ``vZ'' for \cite{vZ},
``B'' for the present work.
The vZ and Sandage numbers were omitted, if not crucial.}
\\[-9pt]
1 & RRab    & $-0.071    $ & $-0.93    $ & $-118.12  $ & R & 	    & \\
2 & not var.        & $1.516     $ & $20.01    $ & $63.28    $ & H & ckn & \\
3 & RRab    & $4.510     $ & $59.52    $ & $-50.41   $ & R &        & \\
4n & RRab         & $-3.006    $ & $-39.68   $ & $1.53     $ & H & d      & GU9003\\
4s & RRab         & $-2.983    $ & $-39.37   $ & $1.16     $ & H & d      & GU9004\\
5 & RRab          & $20.089    $ & $265.13   $ & $-11.14   $ & S &        & \\
6 & RRab          & $-9.098    $ & $-120.07  $ & $69.71    $ & R &        & \\
7 & RRab          & $-0.102    $ & $-1.35    $ & $98.39    $ & H &        & \\
8 & RRab          & $-5.904    $ & $-77.92   $ & $-13.19   $ & R & f      & \\
9 & RRab          & $-21.685   $ & $-286.19  $ & $-198.45  $ & S &        & \\
10 & RRab         & $11.905    $ & $157.11   $ & $148.71   $ & S &        & \\
11 & RRab         & $-11.207   $ & $-147.91  $ & $-199.99  $ & S &        & \\
12 & RRc        & $0.064     $ & $0.85     $ & $-134.77  $ & S &        & \\
13 & RRab         & $-1.611    $ & $-21.27   $ & $-127.31  $ & S &        & \\
14 & RRab         & $-3.381    $ & $-44.62   $ & $-150.70  $ & S &        & \\
15 & RRab         & $-6.519    $ & $-86.04   $ & $-263.30  $ & S &        & \\
16 & RRab         & $-22.480   $ & $-296.69  $ & $-83.69   $ & S &        & \\
17 & RRab         & $11.213    $ & $147.98   $ & $-429.23  $ & S & e      & \\
18 & RRab         & $7.755     $ & $102.36   $ & $-284.70  $ & S &        & \\
19 & RRab         & $26.912    $ & $355.17   $ & $-234.12  $ & S &        & \\
20 & RRab         & $25.626    $ & $338.21   $ & $-260.15  $ & S &        & \\
21 & RRab         & $26.579    $ & $350.79   $ & $29.71    $ & S &        & \\
22 & RRab         & $14.717    $ & $194.23   $ & $0.15     $ & S &        & \\
23 & RRab         & $-8.349    $ & $-110.19  $ & $288.91   $ & S &        & \\
24 & RRab         & $-10.881   $ & $-143.60  $ & $20.11    $ & S &        & \\
25 & RRab         & $-9.111    $ & $-120.24  $ & $-21.66   $ & R &        & \\
26 & RRab         & $-13.149   $ & $-173.54  $ & $-33.40   $ & S &        & \\
27 & RRab         & $-8.019    $ & $-105.83  $ & $-92.68   $ & R &        & \\
28 & RRab         & $-1.566    $ & $-20.66   $ & $-95.40   $ & R &        & \\
29 & RRab         & $-4.608    $ & $-60.81   $ & $-63.42   $ & R & c      & \\
30 & RRab         & $-2.467    $ & $-32.55   $ & $68.13    $ & H &        & \\
31 & RRab         & $2.786     $ & $36.77    $ & $75.62    $ & H &        & \\
32 & RRab         & $1.197     $ & $15.79    $ & $70.53    $ & H &        & \\
33 & RRab         & $5.639     $ & $74.43    $ & $-78.52   $ & R &        & \\
34 & RRab         & $10.523    $ & $138.88   $ & $180.72   $ & S &        & \\
35 & RRab         & $-7.753    $ & $-102.32  $ & $-268.37  $ & S &        & \\
36 & RRab         & $13.344    $ & $176.11   $ & $-24.40   $ & S &        & \\
37 &  RRc         & $-17.633   $ & $-232.72  $ & $173.81   $ & S &        & \\
38 & RRab         & $-15.146   $ & $-199.90  $ & $137.16   $ & S &        & \\
39 & RRab         & $-18.223   $ & $-240.50  $ & $130.63   $ & S &        & \\
40 & RRab         & $-20.257   $ & $-267.35  $ & $121.21   $ & S &        & \\
41 & RRab         & $-6.803    $ & $-89.79   $ & $63.91    $ & R &        & \\
42 & RRab         & $-5.676    $ & $-74.91   $ & $50.59    $ & H &        & \\
43 & RRab         & $7.873     $ & $103.90   $ & $35.09    $ & R &        & \\
44 & RRab         & $13.171    $ & $173.83   $ & $110.21   $ & S &        & \\
45 & RRab         & $-17.959   $ & $-237.02  $ & $-120.53  $ & S &        & \\
46 & RRab         & $-9.373    $ & $-123.71  $ & $-41.30   $ & R &        & \\
47 & RRab         & $-8.578    $ & $-113.21  $ & $-63.05   $ & R &        & \\
48 & RRab         & $9.968     $ & $131.56   $ & $-91.57   $ & R &        & \\
49 & RRab         & $10.915    $ & $144.06   $ & $-89.23   $ & R & e      & \\
50 & RRab         & $1.046     $ & $13.81    $ & $-223.96  $ & S &        & \\
51 & RRab         & $2.677     $ & $35.34    $ & $-215.88  $ & S &        & \\
52 & RRab         & $-5.565    $ & $-73.45   $ & $161.47   $ & S &        & \\
53 & RRab         & $-0.263    $ & $-3.48    $ & $132.94   $ & R &        & \\
54 & RRab         & $-2.207    $ & $-29.12   $ & $116.60   $ & R &        & \\
55 & RRab         & $-15.271   $ & $-201.54  $ & $333.78   $ & S &        & \\
56 &  RRc         & $-10.511   $ & $-138.72  $ & $367.92   $ & S &        & \\
57 & RRab          & $12.060    $ & $159.17   $ & $10.35    $ & S &        & \\
58 & RRab         & $-6.245    $ & $-82.42   $ & $56.12    $ & R &        & \\
59 & RRab         & $-7.951    $ & $-104.94  $ & $-218.58  $ & S &        & \\
60 & RRab         & $-22.135   $ & $-292.14  $ & $-306.30  $ & S &        & \\
61 & RRab         & $14.595    $ & $192.63   $ & $373.81   $ & S &        & \\
62 & RRab         & $7.031     $ & $92.79    $ & $427.27   $ & S &        & \\
63 & RRab         & $3.037     $ & $40.08    $ & $352.17   $ & S &        & \\
64 & RRab         & $8.919     $ & $117.72   $ & $340.62   $ & S &        & \\
65 & RRab         & $9.727     $ & $128.38   $ & $338.11   $ & S &        & \\
66 & RRab         & $-7.409    $ & $-97.78   $ & $131.00   $ & R &        & \\
67 & RRab         & $-9.676    $ & $-127.70  $ & $132.51   $ & R &        & \\
68 & RRd        & $1.907     $ & $25.16    $ & $185.18   $ & S &        & \\
69 & RRab         & $6.379     $ & $84.19    $ & $151.44   $ & R & e      & \\
70 &  RRc         & $3.143     $ & $41.47    $ & $162.56   $ & S &        & \\
71 & RRab         & $12.456    $ & $164.39   $ & $8.52     $ & S &        & \\
72 & RRab         & $34.041    $ & $449.26   $ & $9.47     $ & S &        & \\
73 &   RRc        & $33.506    $ & $442.20   $ & $73.93    $ & S & j      & \\
74 & RRab         & $6.952     $ & $91.75    $ & $161.16   $ & S &        & \\
75 &   RRc        & $3.943     $ & $52.03    $ & $169.47   $ & S &        & \\
76 & RRab         & $-0.779    $ & $-10.28   $ & $-77.78   $ & R &        & \\
77 & RRab         & $-6.902    $ & $-91.10   $ & $37.86    $ & R &        & \\
78 & RRab         & $3.859     $ & $50.93    $ & $76.95    $ & H &        & \\
79 & RRd        & $3.521     $ & $46.47    $ & $359.70   $ & S &        & \\
80 & RRab         & $31.828    $ & $420.06   $ & $295.95   $ & S &        & \\
81 & RRab         & $26.173    $ & $345.43   $ & $362.34   $ & S &        & \\
82 & RRab         & $-7.300    $ & $-96.34   $ & $-591.36  $ & E &        & \\
83 & RRab         & $-33.191   $ & $-438.05  $ & $121.73   $ & S &        & \\
84 & RRab         & $5.123     $ & $67.61    $ & $175.27   $ & S &        & \\
85 &  RRc          & $23.454    $ & $309.54   $ & $237.22   $ & S &        & \\
86 &  RRc        & $39.189    $ & $517.20   $ & $-102.83  $ & S &        & \\
87 &  RRd       & $8.663     $ & $114.34   $ & $70.73    $ & R &        & \\
88 &  RRc        & $-2.355    $ & $-31.07   $ & $-59.58   $ & R &        & \\
89 & RRab         & $2.419     $ & $31.93    $ & $-100.32  $ & R &        & \\
90 & RRab         & $7.725     $ & $101.96   $ & $-177.72  $ & S &        & \\
91 & RRab         & $-0.621    $ & $-8.19    $ & $-539.94  $ & E &        & \\
92 & RRab         & $-1.807    $ & $-23.84   $ & $-398.40  $ & S & e      & \\
93 & RRab         & $-23.787   $ & $-313.94  $ & $-387.62  $ & S &        & \\
94 & RRab         & $-36.648   $ & $-483.67  $ & $-216.49  $ & S &        & \\
95 & long per.        & $-11.427   $ & $-150.81  $ & $25.28    $ & S & & \\
96 & RRab         & $-12.067   $ & $-159.26  $ & $-224.36  $ & S &        & \\
97 &  RRc         & $-9.487    $ & $-125.21  $ & $-187.13  $ & S &        & \\
98 & not var.        & $10.326    $ & $136.28   $ & $7.29     $ & R & ekn    & \\
99 &  RRd       & $15.563    $ & $205.39   $ & $-43.93   $ & S &        & \\
100 & RRab        & $5.566     $ & $73.46    $ & $108.02   $ & R &        & \\
101 & RRab        & $3.795     $ & $50.08    $ & $93.85    $ & R &        & \\
102 &  not var.       & $4.693     $ & $61.94    $ & $125.36   $ & R & kn     & \\
103 &  not var.      & $4.708     $ & $62.13    $ & $130.70   $ & R & kn     & \\
104 & RRab        & $-1.687    $ & $-22.27   $ & $155.46   $ & R &        & \\
105 &  RRc        & $-1.335    $ & $-17.61   $ & $201.42   $ & S &        & \\
106 & RRab        & $-3.393    $ & $-44.78   $ & $178.09   $ & S &        & \\
107 &  RRc        & $-5.551    $ & $-73.26   $ & $344.84   $ & S &        & \\
108 & RRab        & $-16.408   $ & $-216.55  $ & $320.08   $ & S &        & \\
109 & RRab        & $-6.457    $ & $-85.21   $ & $12.52    $ & H &        & \\
110 & RRab        & $-7.259    $ & $-95.81   $ & $-5.71    $ & R &        & \\
111 & RRab        & $-6.735    $ & $-88.89   $ & $31.90    $ & H &        & \\
112 &        & $-10.400   $ & $-137.26  $ & $-709.31  $ & E &        & \\
113 & RRab        & $15.646    $ & $206.49   $ & $-679.26  $ & E &        & \\
114 & RRab        & $1.025     $ & $13.53    $ & $631.56   $ & E &        & \\
115 & RRab        & $33.913    $ & $447.57   $ & $675.49   $ & E &        & \\
116 & RRab        & $-37.143   $ & $-490.20  $ & $473.05   $ & S &        & \\
117 & RRab        & $7.198     $ & $95.00    $ & $-457.98  $ & S & e      & \\
118 & RRab        & $11.306    $ & $149.22   $ & $-281.42  $ & S &        & \\
119 & RRab        & $19.473    $ & $257.01   $ & $117.00   $ & S &        & \\
120 & RRab        & $-22.199   $ & $-292.98  $ & $240.28   $ & S &        & \\
121 & RRab        & $-3.027    $ & $-39.95   $ & $65.91    $ & H &        & \\
122a & RRab       & $-2.148    $ & $-28.35   $ & $-36.74   $ & R & cm     & GU9007\\
122b &       & $-2.231    $ & $-29.45   $ & $-36.36   $ & R & cm     & \\
123  & RRab       & $-18.941   $ & $-249.98  $ & $-988.00  $ & E &        & \\
124  & RRab       & $-4.671    $ & $-61.65   $ & $-191.15  $ & S &        & \\
125  &  RRc       & $14.449    $ & $190.69   $ & $-121.86  $ & S &        & \\
126  &  RRc       & $-0.837    $ & $-11.04   $ & $-136.10  $ & S &        & \\
127  &            & $7.364     $ & $97.19    $ & $-50.42   $ & R & cn     & \\
128  &   RRc      & $8.926     $ & $117.80   $ & $141.70   $ & R &        & \\
129  &  RRc       & $-2.986    $ & $-39.41   $ & $87.55    $ & H &        & \\
130  & RRab       & $0.577     $ & $7.61     $ & $94.30    $ & H &        & \\
131  &  RRc       & $-5.289    $ & $-69.81   $ & $37.27    $ & H &        & \\
132  &  RRc       & $-3.757    $ & $-49.58   $ & $-11.85   $ & H &        & GU9001\\
133  & RRab       & $-4.176    $ & $-55.12   $ & $54.18    $ & H &        & \\
134  & RRab       & $-1.409    $ & $-18.60   $ & $62.87    $ & H &        & \\
135  & RRab       & $-1.747    $ & $-23.06   $ & $48.69    $ & H &        & \\
136  & RRab       & $-1.625    $ & $-21.44   $ & $44.33    $ & H & f      & \\
137  & RRab       & $4.275     $ & $56.42    $ & $-8.71    $ & H &        & \\
138  & long per.      & $-19.662   $ & $-259.49  $ & $50.72    $ & S &        & \\
139  & RRab       & $2.916     $ & $38.49    $ & $38.60    $ & H &        & \\
140  &  RRc       & $-0.901    $ & $-11.90   $ & $119.17   $ & H &        & \\
141  &  EW     & $-113.050  $ & $-1492.00 $ & $-250.67  $ & E &        & RV CVn\\
142  & RRab       & $-1.934    $ & $-25.53   $ & $-48.12   $ & R &        & Sh2, GU9009\\
143  & RRab       & $-2.294    $ & $-30.27   $ & $27.22    $ & H &        & GU9005\\
144  & RRab       & $4.406     $ & $58.15    $ & $-89.10   $ & R &        & \\
145  & RRab       & $2.421     $ & $31.95    $ & $19.18    $ & H & m      & \\
146  & RRab       & $7.301     $ & $96.37    $ & $-47.91   $ & R & c      & \\
147  &  RRc       & $-1.364    $ & $-18.00   $ & $57.57    $ & H &        & \\
148  & RRab       & $-0.239    $ & $-3.16    $ & $48.21    $ & H &        & \\
149  & RRab       & $2.903     $ & $38.32    $ & $63.43    $ & H &        & \\
150  & RRab       & $5.487     $ & $72.42    $ & $48.61    $ & H &        & \\
151  & RRab       & $0.789     $ & $10.42    $ & $-30.86   $ & R & c      & \\
152  &  RRc       & $6.269     $ & $82.74    $ & $61.77    $ & R &        & \\
153  & not var.   & $-2.585    $ & $-34.12   $ & $71.71    $ & H &  kn & \\
154  & W Vir      & $0.463     $ & $6.10     $ & $-18.26   $ & H & cm     & GU9024\\
155  & RRc      & $-4.500    $ & $-59.39   $ & $-64.74   $ & R & c      & Sh1\\
156  & RRab       & $-1.212    $ & $-15.99   $ & $-30.39   $ & R & c      & GU9019?\\
157  & RRab       & $-1.013    $ & $-13.36   $ & $46.13    $ & H &        & \\
158  & not var.      & $-0.857    $ & $-11.30   $ & $-29.49   $ & R & ckn    & \\
159a & RRab       & $-0.842    $ & $-11.12   $ & $26.90    $ & H & cd     & \\
159b &       & $-0.833    $ & $-11.00   $ & $26.67    $ & H & cd     & \\
160  & RRab       & $-0.394    $ & $-5.20    $ & $-32.75   $ & R &        & GU9021\\
161  & RRab       & $1.611     $ & $21.27    $ & $-46.98   $ & R & c      & \\
162  & not var.      & $2.404     $ & $31.73    $ & $-20.75   $ & R & kn     & \\
163  & not var.      & $-0.903    $ & $-11.92   $ & $-21.25   $ & R & cn     & \\
164  &       & $1.890     $ & $24.94    $ & $-25.18   $ & R & j?      & \\
165  & RRab       & $5.846     $ & $77.16    $ & $31.01    $ & H & c      & \\
166  & RRd      & $-7.031    $ & $-92.79   $ & $2.69     $ & R &        & \\
167  & RRab       & $-5.613    $ & $-74.08   $ & $-26.25   $ & R &        & \\
168  &  RRc       & $-3.121    $ & $-41.18   $ & $17.65    $ & H &        & GU9002\\
169  &  not var.      & $-1.918    $ & $-25.31   $ & $-23.85   $ & R & kn     & \\
170  &  RRc       & $-1.863    $ & $-24.59   $ & $43.45    $ & H & c      & \\
171  &  RRc       & $-1.686    $ & $-22.25   $ & $27.40    $ & H &        & Sh3\\
172  & RRab       & $-1.293    $ & $-17.07   $ & $36.67    $ & H &        & \\
173  & RRab       & $-0.691    $ & $-9.13    $ & $49.66    $ & H &        & \\
174  & RRab       & $-0.367    $ & $-4.85    $ & $-23.30   $ & R &        & GU9022\\
175  & RRab       & $3.437     $ & $45.35    $ & $37.17    $ & H &        & \\
176  & RRab       & $3.783     $ & $49.93    $ & $43.95    $ & H &        & \\
177  &  RRc       & $5.068     $ & $66.89    $ & $-18.21   $ & R &        & \\
178  &  RRc       & $6.279     $ & $82.86    $ & $57.72    $ & R &        & \\
179  &       & ?            & ?           & ?           &  &  ci      & \\
180  & RRab      & $-1.111    $ & $-14.66   $ & $-18.53   $ & R & cfi    & Sh5?\\
181  & RRab       & $-2.001    $ & $-26.40   $ & $-2.98    $ & H & cf     & GU9008\\
182  &  not var.      & $-1.076    $ & $-14.20   $ & $70.47    $ & H & kn & \\
183  &  not var.      & $2.580     $ & $34.05    $ & $17.63    $ & H & kn     & \\
184  & RRab       & $-1.624    $ & $-21.43   $ & $-4.35    $ & R &        & GU9013\\
185  &            & $-0.883    $ & $-11.66   $ & $41.84    $ & H & cn     & \\
186  & RRab       & $1.301     $ & $17.18    $ & $-52.32   $ & R &        & \\
187  & RRab     & $-1.550    $ & $-20.46   $ & $19.82    $ & H & m      & \\
188  &  RRc       & $-1.704    $ & $-22.49   $ & $34.72    $ & H &        & \\
189  & RRab      & $-1.611    $ & $-21.27   $ & $-10.18   $ & R &        & Sh4, GU9014\\
190  & RRab      & $-0.319    $ & $-4.22    $ & $39.16    $ & H &        & \\
191  & RRab      & $0.426     $ & $5.62     $ & $33.81    $ & H & m      & \\
192a & RRab      & $0.112     $ & $1.48     $ & $14.87    $ & H & cd     & \\
192b &       & $0.112     $ & $1.48     $ & $14.74    $ & H & cd     & \\
193  & RRab      & $1.401     $ & $18.50    $ & $3.63     $ & H &        & GU2538\\
194  & RRab      & $1.550     $ & $20.45    $ & $-1.98    $ & H & cm    & KG10\\
195  & RRab       & $-0.708    $ & $-9.35    $ & $-17.29   $ & R &        & GU9020\\
196a &  not var.      & $3.848     $ & $50.79    $ & $11.60    $ & H & cikn   & \\
196b &  not var.      & $3.811     $ & $50.30    $ & $12.22    $ & H & cikn   & \\
196c &  not var.     & $3.915     $ & $51.67    $ & $13.18    $ & H & cikn   & \\
197  & RRab       & $4.713     $ & $62.19    $ & $20.03    $ & H &        & \\
198  &         & $-1.449    $ & $-19.13   $ & $25.74    $ & H & cn     & \\
199  & RR      & $-1.168    $ & $-15.42   $ & $21.92    $ & H &    & \\
200  & RRab      & $-0.021    $ & $-0.28    $ & $31.95    $ & H & cfm    & \\
201  & RRab       & $0.585     $ & $7.73     $ & $1.82     $ & H &        & GU1600\\
202  &  RRc       & $-28.355   $ & $-374.23  $ & $105.76   $ & S &        & \\
203  &  RRc       & $-1.751    $ & $-23.11   $ & $-301.59  $ & S &        & \\
204  &  not var.      & $-7.583    $ & $-100.09  $ & $-12.08   $ & R & ckn    & \\
205  & RRab       & $-83.596   $ & $-1103.28 $ & $653.86   $ & E &        & WY CVn\\
206  & RRab       & $9.113     $ & $120.27   $ & $-1720.86 $ & E &        & WZ CVn\\
207  &  RRc       & $3.017     $ & $39.82    $ & $-20.66   $ & R & f      & X3\\
208  &  RRc       & $0.509     $ & $6.71     $ & $-47.73   $ & R & f      & X4\\
209  &  RRc       & $-4.836    $ & $-63.82   $ & $-89.17   $ & R & cf     & X6\\
210  & RRc      & $-6.213    $ & $-82.00   $ & $0.35     $ & H & fm     & X8\\
211  & RRab       & $-3.797    $ & $-50.11   $ & $17.17    $ & H & cd     & X9\\
212  & RRab       & $-1.329    $ & $-17.53   $ & $-27.78   $ & R &        & X11, GU9017\\
213  &  RRc       & $-1.630    $ & $-21.51   $ & $-19.40   $ & R &        & X12, GU9011\\
214  & RRab       & $2.734     $ & $36.08    $ & $16.70    $ & H &        & X15\\
215  & RRab       & $-0.759    $ & $-10.02   $ & $9.52     $ & H &        & X16, GU238\\
216  &  RRc       & $2.399     $ & $31.67    $ & $-0.59    $ & H & f      & X19\\
217  & RRab       & $0.278     $ & $3.67     $ & $-16.43   $ & H &        & X27, GU9023\\
218  & RRab       & $2.424     $ & $31.99    $ & $-19.01   $ & R &        & X28\\
219  &  RRab       & $-4.139    $ & $-54.62   $ & $25.72    $ & H & fm     & X29\\
220  & RRab       & $2.801     $ & $36.97    $ & $-5.50    $ & H & f      & X31\\
221  &  RRc       & $-0.985    $ & $-12.99   $ & $-3.19    $ & H &        & X32, GU85\\
222  & RRab       & $7.565     $ & $99.84    $ & $-52.96   $ & R &        & X34\\
223  &  RRc       & $2.083     $ & $27.49    $ & $4.46     $ & H & f      & X37\\
224  &       & $-1.375    $ & $-18.15   $ & $15.16    $ & H & f?m      & X38\\
225  & long per.      & $0.863     $ & $11.38    $ & $234.63   $ & S &        & \\
226  & RRab       & $1.783     $ & $23.54    $ & $-8.04    $ & R &        & X39\\
227  &  not var.      & $-8.048    $ & $-106.21  $ & $-51.40   $ & R & kn     & X7\\
228  &  not var.     & $-2.229    $ & $-29.41   $ & $46.80    $ & H & kn     & X10\\
229  & RRab       & $-2.194    $ & $-28.95   $ & $-35.13   $ & R & cm     & X40, GU9006\\
230  &       & $19.733    $ & $260.44   $ & $-482.84  $ & E &        & X41\\
231  &  not var.      & $9.731     $ & $128.42   $ & $57.73    $ & R & kn     & X42\\
232  &  not var.     & $9.310     $ & $122.87   $ & $52.22    $ & R & kn     & X43\\
233  &            & $2.618     $ & $34.55    $ & $3.03     $ & H & n      & X44\\
234  & RRab       & $1.889     $ & $24.93    $ & $-29.30   $ & R & f      & X47\\
235  & RRab       & $2.573     $ & $33.95    $ & $47.61    $ & H & f      & X49\\
236  & long per.      & $30.337    $ & $400.38   $ & $-203.99  $ & S & j      & vZ1397\\
237  & SX Phe      & $1.757     $ & $23.19    $ & $-170.11  $ & S & c      & \\
238  & EW      & $4.551     $ & $60.06    $ & $-254.85  $ & S & c      & \\
239  & RRab       &$-1.326    $ & $-17.50   $ & $-16.28   $ & R & ci     & X13, GU9018, Sh5?\\
240  &  RRc &	$-1.739    $ & $-22.95   $ & $3.10     $ & H & cd     & X14a, GU9010\\
     &  & $-1.713    $ & $-22.61   $ & $3.00     $ & H & cd     & X14b, GU9010?\\
241  & RRab       &  $-0.435    $ & $-5.74    $ & $5.78     $ & H &  cm      & X17, GU507\\
     &  	& $-0.326    $ & $-4.30    $ & $2.23     $ & H & cn     & X18\\
242  & RRab		  & $2.009     $ & $26.52    $ & $-7.63    $ & R & cfm    & X20 \\
243  & RRab	& $1.071     $ & $14.13    $ & $-16.33   $ & R &        & X22, GU9026\\
244  & RRab		  & $-1.575    $ & $-20.78   $ & $14.41    $ & H & cf     & X23 \\
245  & RRc		  & $-1.300    $ & $-17.16   $ & $27.85    $ & H & f      & X25 \\
246  & RRc           & $1.562     $ & $20.62    $ & $8.33     $ & H & f      & X30 \\
247  &  RRab	   & $3.593     $ & $47.43    $ & $-16.32   $ & R & df     & X35 \\
248  &	RRab	   & $-1.532    $ & $-20.22   $ & $17.52    $ & H & c      & X36, GU9015\\
249  &	RRab	   & $-0.887    $ & $-11.71   $ & $16.07    $ & R & cf & KG1 \\
250  &	RRab	   & $-0.642    $ & $-8.48    $ & $19.92    $ & H & cdf    & KG2 \\
251  &  RR   & $-0.245 $ & $-3.23$  & $1.79 $ & H & cim       & KG3, GU684\\
252  &	RRab	   & $-0.202    $ & $-2.67    $ & $11.79    $ & H &        & KG4, GU734\\
     &   	     & $0.504     $ & $6.65     $ & $-6.38    $ & R & cn     &  KG5 \\
     &   	     & $0.871     $ & $11.49    $ & $13.66    $ & H & kn      & KG6 \\
253  &	RRc     & $0.942     $ & $12.44    $ & $0.62     $ & H &        & KG7, GU2042\\
254  &	RRab     & $1.217     $ & $16.06    $ & $21.39    $ & H & cdfm   & KG8a \\
     &       & $1.258     $ & $16.60    $ & $21.03    $ & H & cdfm   & KG8b \\
255  &	RR     & $1.409     $ & $18.59    $ & $11.81    $ & H & f      & KG9 \\
256  &	RRc     & $1.896     $ & $25.03    $ & $26.58    $ & H & cdf    & KG11 \\
257  &  RRab     & $1.857     $ & $24.50    $ & $-3.80    $ & R & f      & KG12 \\
     &           & $2.335     $ & $30.82    $ & $-30.71   $ & R & cin    & KG13 \\
258  &	RR	  & $3.112     $ & $41.07    $ & $59.38    $ & H & f      & KG14 \\
259  &	RRc	  & $3.373     $ & $44.51    $ & $22.70    $ & H & cdfj    & KG15 \\
     &            & $-6.320    $ & $-83.41   $ & $-88.45   $ & R & k      & KG16 \\
     &            & $6.080     $ & $80.24    $ & $265.22   $ & R & k     & KG17, S-I-II-53\\
	&       & $-34.879   $ & $-460.32  $ & $-67.45   $ & S & j?     & vZ177 \\
	&       & $-26.927   $ & $-355.38  $ & $212.99   $ & S & k      & vZ194 \\
	&       & $-18.005   $ & $-237.62  $ & $-186.21  $ & S & j?     & vZ250 \\
	&       & $-16.158   $ & $-213.25  $ & $-196.90  $ & S & k      & vZ265\\
	&       & $-14.865   $ & $-196.19  $ & $204.76   $ & S & j?     & vZ281 \\
260     & long per.	& $-13.023   $ & $-171.88  $ & $-399.92  $ & S & j      & vZ297 \\
	&      & $-7.963    $ & $-105.10  $ & $41.71    $ & R & ckn    & vZ380\\
	&      & $1.175     $ & $15.50    $ & $-234.43  $ & S & j?     & vZ853 \\
	&      & $4.137     $ & $54.59    $ & $97.54    $ & S & cn     & vZ1066 \\
	&      & $11.379    $ & $150.18   $ & $-45.84   $ & R & ckn    & vZ1283\\
	&      & $17.607    $ & $232.37   $ & $104.86   $ & S & j?     & vZ1345\\
	&      & $20.205    $ & $266.67   $ & $22.55    $ & S & j?     & vZ1360 \\
	&      & $-11.823   $ & $-156.04  $ & $217.50   $ & S & k & S-I-III-38\\
	&      & $-12.881   $ & $-170.00  $ & $-302.20  $ & S & k      & S-O\\
	&      & $-9.852    $ & $-130.03  $ & $209.23   $ & S & k      & S-AK \\
	&      & $14.830    $ & $195.72   $ & $268.75   $ & S & k      & S-I-I-27\\
	&   & $12.720    $ & $167.87   $ & $-77.30   $ & S & j?     & S-I-VI-65 \\
	&   & $3.878     $ & $51.18    $ & $273.98   $ & S & ck     & S-AQ  \\
	&   & $5.147     $ & $67.92    $ & $267.31   $ & S & ck     &  S-I-II-52 \\
	&   & $5.910     $ & $78.00    $ & $261.32   $ & S & ck     & S-I-II-54\\
261	& RRc   & $-1.113    $ & $-14.69   $ & $8.18     $ & H &        & GU32\\
262 	& RR      & $-0.384    $ & $-5.07    $ & $5.28     $ & H & c      & GU552\\
263     & SX Phe   & $-0.329    $ & $-4.35    $ & $-7.82    $ & H & c & GU576\\
264 	& RRc   & $-0.329    $ & $-4.35    $ & $-2.31    $ & R &        & GU586\\
265     & RRab    & $0.512     $ & $6.76     $ & $-16.36   $ & H & cm     & GU1489\\
	&     & $0.673     $ & $8.87     $ & $5.73     $ & H & cn     & GU1711\\
266  	& RRc    & $-1.648    $ & $-21.75   $ & $12.14    $ & H &        & GU9012\\
267  	& SX Phe    & $-1.416    $ & $-18.69   $ & $15.87    $ & H & cm     & GU9016\\
268  	& RRab    & $0.512     $ & $6.76     $ & $-17.80   $ & H & cm     & GU9025\\
269  & RRc   & $1.590     $ & $20.98    $ & $0.88     $ & R & m      & B1\\
270  & RRab	    & $0.746     $ & $9.85     $ & $59.97    $ & R & m      & B2\\
271  & RRab	    & $0.992     $ & $13.09    $ & $45.84    $ & R & m      & B3\\
272  & long per.	    & $2.087     $ & $27.54    $ & $-161.58  $ & S & & B4\\
273  & long per.	    & $-4.469    $ & $-58.98   $ & $-95.07   $ & S & & B5\\
274  & long per.	    & $0.901     $ & $11.89    $ & $-71.47   $ & S & & B6\\

\end{longtable}
\normalsize

\appendix

\section{Notes on individual stars}
\footnotesize
\noindent {\bf V2:}
Two closeby candidates in \cite{ESTS94} (hereafter ESTS94); none of them is
variable. From charts in \cite{Kholopov1963,Kholopov1977} (hereafter K63
and K77) it is not obvious which star is marked as ``2''. Original positions
from \cite{SH73} (hereafter SH73) and \cite{CC98} (hereafter CC98) are
closer to the west candidate.

\noindent {\bf V29:}
Positions of V29 and V155 are interchanged in ESTS94.

\noindent {\bf V122:}
Two candidates; V122a is more likely the true variable,
based upon its light-curve. Merging with V229. 

\noindent {\bf V127:}
CC98 position merges with position of V222. K77 gives correct 
identification.

\noindent {\bf V146:}
CC98 position is $\sim4''$ away from the variable. As V127 and V222 are
closeby, this position error might be misleading. Identification confirmed
from K77 charts.

\noindent {\bf V151:}
Quite far ($\sim3''$) from the CC98 position, the same star is marked in
K77. No other candidate.

\noindent {\bf V154:}
W Vir star, merging with an RR Lyrae V268.

\noindent {\bf V155:}
Positions of V29 and V155 are interchanged in ESTS94.

\noindent {\bf V156:}
Formal position of GU9019 is $1.3''$ away from V156, not coinciding
with any
star. It is likely to be the same variable as V156.

\noindent {\bf V158:}
Despite the entry in CC98, it doesn't seem to be variable on the RCC images
(one night), neither on the Schmidt images. 

\noindent {\bf V159:}
Double on HST/WF2 images: V159a, V159b. Not possible to select the variable
component. 

\noindent {\bf V161:}
Misidentified in ESTS94 (with non-variable star in $\sim2''$ distance). 

\noindent {\bf V163:}
Star from \cite{Muller1933}; identification is dubious, position is equidistant from
a non-variable and a variable star (later V180). K63 and K77 marks the
non-variable candidate as 163. CC98 says V163 is not var. The not variable
candidate is accepted as V163, in order to avoid confusion.

\noindent {\bf V165:}
The coordinates in SH73, CC98 are erroneous (sign on declination offset is
``$-$'' instead of ``+'')

\noindent {\bf V170:}
Elongated profile on HST/WF2 image. Maybe double?

\noindent {\bf V179:}
This candidate is very far from the cluster, only crude position 
is known, charts are not available. Three
stars are 
equidistant on the POSS images from the quoted position.
CC98 notes that this star is non-variable.

\noindent {\bf V180:}
Two equidistant candidates close to SH73 (Sh) position:
one is the same as X13=(V239), which is variable indeed. K77 marks the other star as
180, and announces X13 as a new variable. Shapley says that V180 is non-variable.
For convenience, the candidate shown in K77 is chosen to be V180. Variability
is confirmed, as opposed to CC98. 

\noindent {\bf V181:}
Appropriately transformed CC98 position coincides with a variable star.
However, CC98 notes that this star is not variable. K77 finder charts mark a
closeby, non-variable star as 181. ESTS94 identifies the same star as V181. 
We propose that both K77 and ESTS94 identifications are erroneous, and V181
is variable, identical to GU9008. Faint companion to the NW on HST/WF2
image.

\noindent{\bf V185:}
No variable star in the vicinity of the position in CC98. V185, as
identified in ESTS94 shows no variability.

\noindent {\bf V192:}
Obviously double on HST/PC image; not possible to select the variable. 

\noindent {\bf V194:}
Two candidates on HST/PC frame. Merging with other stars to the N in a NS
chain, all of them showing variability? Time-series observations with better
seeing would be needed for firm identification. Formal position of KG10, as
derived from \cite{KG80} (hereafter KG80) and ESTS94, is merging with
V194. 

\noindent {\bf V196:}
CC98 position is close to a constant star, but it is not the same as the one
marked in K77 finder charts. The latter is also constant (196c). The star
close to the CC98 position, is a double star on the HST/WF4 image (196a,
196b), both components are constant.

\noindent {\bf V198:}
Despite the entry in CC98 and the short period ($\sim 0.2797^d$), it doesn't
seem to be variable on the RCC images (one night).

\noindent {\bf V200:}
Misidentified in ESTS94, merging with a brighter star. 

\noindent {\bf V204:}
Roughly $6''$ away from the position in CC98; identification from
K77 chart. No significant variability detected on the RCC frames (one
night), neither on the Schmidt frames. 

\noindent {\bf V209:}
ESTS94 lists two stars at the position. Unfortunately the area is not
within the HST fields. Variable is more likely to be the southern one
(from variability image).

\noindent {\bf V211:}
Obviously double on HST images. 

\noindent {\bf V229:}
Merging with V122a and V122b. 

\noindent {\bf V237, V238:}
Too faint stars for variability to be confirmed on Schmidt images. 

\noindent {\bf V239=X13:} 
Confer comments at V180.

\noindent {\bf V240=X14:}
Obviously double on HST/WF2 images: X14a, X14b. Maybe both components are
variable, as the source is elongated on the variability image.

\noindent {\bf V241=X17:}
See comment at V262.

\noindent {\bf X18:}
K77 marks a relatively bright star as X18, which shows no significant
variability on the RCC frames spanning one day. Very close to formal
position of V251 (=KG3).

\noindent {\bf V242=X20:}
Merging with a bright star, and close to V257. 

\noindent {\bf V244=X23:}
Two candidates on HST/WF2 image: the true variable was selected using
the RCC variability image. X23 is not identical with GU9012, as
proposed by \cite{Goranskij1994}.

\noindent {\bf V248=X36:}
Elongated profile on HST/WF2 image.

\noindent {\bf V249=KG1:}
Identification with the help of KG80 positions adequately
transformed to our system.

\noindent {\bf V250=KG2:}
Two close candidates on HST/WF2 image. 

\noindent {\bf V251=KG3:}
Merging with X18, only resolved on HST/PC frames. Variability (due to
the extreme crowding) not detected unambigously.

\noindent {\bf KG5:}
Bright star, identified using both ESTS94 and KG80: no significant
variability.

\noindent {\bf V254=KG8:}
Two candidates on HST/WF4 image, merging on gound-based images. 
Light curves are almost the same for the two slightly different positions. 
Further investigation in good seeing would be needed to draw firm
conclusion.

\noindent {\bf V256=KG11:}
Two candidates on HST/WF4 images; the true variable was selected using
the RCC variability image.

\noindent {\bf KG13:}
Identification dubious; two bright stars and a fainter companion close to   
the position showing no significant variability.

\noindent {\bf V259=KG15:}
Two candidates on HST/WF4 images; the true variable was selected using
the RCC variability image.

\noindent {\bf vZ380:}
Star in ESTS94, variability from \cite{Meinunger1980} \--- obj. a.

\noindent {\bf vZ1066:}
Star in ESTS94, variability from \cite{Meinunger1980} \--- obj. b.

\noindent {\bf vZ1283:}
Star in ESTS94, variability from \cite{Meinunger1980} \--- obj. c.

\noindent {\bf S-AQ, S-I-II-52, S-I-II-54:}
Low amplitude variability
\cite{Bao-An1993a}, \cite{Bao-An1993b}, \cite{Bao-An1994} is
not detected on the Schmidt images.

\noindent {\bf V262=GU552:}
This might be a very close companion to V241, only resolved on HST
images. Light curve is almost identical to that of the closeby V241, due to merging.
No trace of RRc variation is detectable in the light-curve, in spite of
comments in \cite{GU94}. Higher resolution time-series observations would be needed
to conclude more on its variability.

\noindent {\bf V263=GU576:}
SX Phe, barely visible on the 1m RCC images. Position from HST data.
Elongated on HST/WF2?

\noindent {\bf V265=GU1489:}
RR Lyrae, merging with V154 and V268. 

\noindent {\bf GU1711:}
Suspected variable in \cite{GU94}, constant on the RCC images. 

\noindent {\bf V267=GU9016:}
SX Phe star merging with V224, resolved on the HST/WF2 frame. 

\noindent {\bf V268=GU9025:}
RR Lyrae star merging with V154 (W Vir) and V265.


\end{document}